# Sea Quark and Gluon contributions to Strange Baryons and their Properties

M. Batra, A. Upadhyay
School of Physics and Material Science
Thapar University, Patiala, Punjab, 147004

**Abstract:** We treat the hadrons as an ensemble of quark-gluon Fock states where contributions from sea-quarks and gluons can be studied in detail for the properties of low lying baryons. Statistical model is applied to calculate individual probabilities from various scalar, vector and tensor sea components in flavor, spin and color subspaces for each quark-gluon Fock state. The mass of strange quark is imposed in terms of constraints to the availability of strange quarks in the sea and free energy of gluons in conformity with experimental indications. We calculate multiplicities for different sets of Fock states to compute the role of strange sea for cascade doublet $\Xi^+$, and $\Xi^-$. The low energy properties like magnetic moment, weak decay matrix elements and axial coupling constant ratios have been studied. The incorporation of strange quark gluon sea is discussed and they found to affect the results by almost 46% . The results provide deeper understanding for baryon structure thereby motivating experiments for further inspection, especially the spin distribution among the quarks and gluons.

## I Introduction:

In the low energy limit, QCD has a challenging behavior. The non-perturbative and relativistic nature of quarks and gluons makes the hadronic structures difficult to understand. The non-perturbative mechanism can be made to understand via several models. Although the constituent quark model tends to explain the static properties of a baryonic system, still there is a need of some new degrees of freedom. The extension of valence quarks to sea with quark-antiquark pair is a remarkable step in hadron spectroscopy. A variety of approaches have been developed to understand the behavior of baryons with three valence quarks and sea consisting of gluons and quark-antiquark pair. These models help to make a deeper understanding to the various properties like weak decay coupling ratios, magnetic moments, masses of hadrons and flavor asymmetry. Latest studies, dedicated towards the study of quark-antiquark pairs and gluon condensates includes statistical balance model[1], lattice QCD[3], effective QCD approaches[2] and meson-cloud based approaches[4-6]. The static properties of hadrons had also been studied through a large variety of experiments [7-9]. Recent experimental studies show that strange quark also make a significant contribution to nucleonic spin which can be studied by using polarized deep-inelastic scattering experiments of electrons or muons from nucleons. The aim of present work is to analyze the probabilities of various Fock states in different cases. The sea is assumed to contain strange as well as non-strange quark and their contributions to various properties are analyzed by determining the multiplicities in color, flavor and spin states. The multiplicities are used to obtain the probabilities for each quark-gluon Fock states in the ground state baryons. In section II, we briefly present the theoretical frame. Numerical results are given in section III, putting forward predictions for the probabilities of all Fock states for Cascade baryonic system.

## II Theory:

A wave-function ( $\Psi = \Phi(|\phi\rangle|\chi\rangle|\psi\rangle|\xi\rangle)$ ) including 3-quark wave-function and suitable sea contents is framed keeping in mind the total anti-symmetry of the wave-function where $|\phi\rangle, |\chi\rangle, |\psi\rangle, |\xi\rangle$ is for flavor, spin and color and space-time wave-function respectively[13]. Let $\varphi_1^{1/2}$ is the standard SU(3) $q^3$ wave-function transforming as 56 of SU(6) wave-function. $H_{0,1,2}$ and $G_{1,8,10}$ denotes the spin and color possibilities of sea with spin and color combinations of 0,1,2 and 1,8,10 respectively. The total flavor-spin-color wave function of a spin up baryon which consists of three-valence quarks and sea components can be written as given below[13]. All the terms have to be written properly with appropriate CG coefficients and by taking into account the symmetry property of the component wave function. To obtain all the coefficients in the wave-function, a general parameterization method is employed to calculate the different observables in the form of eigenvalues for baryonic system.

$$\left|\Phi^{\uparrow}_{\frac{1}{2}}\right\rangle = \frac{1}{N}[\phi_1^{(\frac{1}{2})\uparrow} H_0 G_1 + a_8 \phi_8^{(\frac{1}{2})\uparrow} H_0 G_8 + a_{10} \phi_{10}^{(\frac{1}{2})\uparrow} H_0 G_{\overline{10}}$$

$$+ b_1 [\phi_1^{\frac{1}{2}} \otimes H_1]^{\uparrow} G_1 + b_8 (\phi_8^{\frac{1}{2}} \otimes H_1)^{\uparrow} G_8 +$$

$$b_{10}(\phi_{\overline{10}}^{\frac{1}{2}} \otimes H_1)^{\uparrow} G_{\overline{10}} + c_8 (\phi_8^{\frac{3}{2}} \otimes H_1)^{\uparrow} G_8 +$$

$$d_8 (\phi_8^{\frac{3}{2}} \otimes H_2)^{\uparrow} G_8]$$

Where

$$N^2 = 1 + a_8^2 + a_{10}^2 + b_1^2 + b_8^2 + b_{10}^2 + c_8^2 + d_8^2$$

Assuming the expansion of hadrons in terms of quark-gluon Fock states where Fock states avail the presence of quark-antiquark pairs multi-connected to gluons. Mathematically, it can be expressed as:

$$|h\rangle = \sum_{i,j,k} c_{i,j,l,k} |\{q\},\{i,j,l\},\{k\}\rangle$$

where $\{q\}$ represents the valence quarks of the baryon, i,j,l represents the number of $q\bar{q}(u\bar{u},d\bar{d},s\bar{s})$ and k is the number of gluons. The principle states that the transition between any two processes either via splitting or recombination involving any two kind of partons should balance each other. The splitting and recombination procedure involves three kinds of processes and their transition probabilities involve number of gluons and quark-antiquark pairs [10-11].

(i) When $q \Leftrightarrow qg$ is considered: The general expression of probability for these sub-processes can be described as:
$$\frac{\rho_{i,j,l,k}}{\rho_{i,j,l,k-1}} = \frac{1}{k}$$

(ii) When both the processes $g \Leftrightarrow gg$ and $q \Leftrightarrow qg$ are included: we can write

$$\frac{\rho_{i,j,l,k}}{\rho_{i,j,l,k-1}} = \frac{(3+2i+2j+2l+k-1)}{(3+2i+2j+2l)k + \frac{k(k-1)}{2}}$$

(iii) When the processes $g \Leftrightarrow q\bar{q}$ are involved: The transition probabilities involving $g \Leftrightarrow q\bar{q}$ depend upon the valence quark content and differ in all baryons. The generation of $s\bar{s}$ pair from gluons is restricted due to non-negligible mass of s-quark. The constraint on the free energy of gluon comes in the form of factor $k(1-C_l)^{n-1}$ [12] where k is the number of gluons and n represents the total number of partons present in that state. Therefore, the splitting and recombination for the processes involving $g \Leftrightarrow q\bar{q}$ undergoes SU(3) symmetric breaking in sea where q is for some heavier quark flavor.

In general, $n = 3 + 2i + 2j + 2l + k$

and $C_{l-1} = \frac{2M_s}{M_B - sM_s - 2(l-1)M_s}$, $M_s$ is the mass of s-quark and $M_B$ is the mass of the baryon.

$$|\{q\},i,j,l+k-1,1\rangle \xrightarrow[l+k(l+k+1)]{1(1-C_{l-1})^{n-k-2}} |\{q\},i,j,l+k,0\rangle$$

$$\frac{\rho_{i,j,l,k}}{\rho_{i,j,l+k,0}} = \frac{(k(k-1)--1(1-C_0)^{n-2l-1}--(1-C_{l-1})^{n+k-2})}{(l+1)(l+2)\ldots\ldots(l+k)(l+k+1)}$$

The above expression produces all the probabilities in the form of ratios. All the probabilities in flavor space can be obtained from the table given below for individual particle. Complete expressions for cascade doublet can be written as:

| Baryon | Expression |
|---|---|
| $\Xi^0$ uss | $\frac{\rho_{i,j,l+k,0}}{\rho_{0,0,0,0}} = \frac{2}{i!\,i+1!\,j!(j)!(l+k)!(l+k+2)!}$ |
| $\Xi^-$ dss | $\frac{\rho_{i,j,l+k,0}}{\rho_{0,0,0,0}} = \frac{2}{i!\,i!\,j!(j+1)!(l+k)!(l+k+2)!}$ |

**Table 1 Probability Expressions**

These probabilities in flavor space is further needed to calculate probabilities in spin and color space [14-15]. The decomposition of baryonic states in various quark-gluon Fock states with relevant operators acting on the valence and sea part is used to find probabilities for all possible spin and color sub-states so as to produce spin ½ and color singlet state of the baryon. The procedure can be applied to states like $|\bar{u}ug\rangle, |\bar{d}dg\rangle, |\bar{s}sg\rangle$, $|\bar{d}d\bar{s}s\rangle, |gg\rangle$ and $|\bar{u}u\bar{d}d\rangle$. More favorable situation occurs when the higher multiplicities are suppressed with the fact that the larger multiplicity leads to higher possibility of interaction and less survival probability. Moreover, non-zero mass of s-quark limits the free energy of gluon and hence assumed to be less probable. The suppression of states with higher multiplicities can be made to archive on phenomenological grounds by assuming the probability of a system being in a particular spin and color state is inversely proportional to multiplicity of the spin and color of that state. The probability in terms of multiplicity for light and strange quarks are mentioned in tables 2 and 3.The sum of all probabilities in individual cases provide a suitable normalization constant and the coefficients in the wave-function for baryon. Static properties of ground state strange baryon octets can be determined from these coefficients.

**III Results and Discussions:**

We aim to choose one set of the strange baryon i.e $\Xi^-, \Sigma^0, \Sigma^+$ a isospin triplet to compute probabilities of several Fock states in spin, flavor and color space and to check the validity of the statistical model to study their low energy properties. The study has been divided into three cases to study the dominant role of strange sea with proper assumptions. Case I comes with the assumption that gluon has a possibility to create $q\bar{q}$ pair (including strange quark) which carries the quantum numbers of gluon, hence it suggests the suppressed multiplicity of $q\bar{q}$ and g to have spin 1, In case II we rule out the assumption for $q\bar{q}$ pair to have the same quantum numbers as that of gluon, and hence it can be in spin state 0,1, or 2. Hence enhanced possibility through which it can contribute to spin 1/2 baryon. Third case is similar to the first case where we modify the detailed balance principle by ruling out the sub-processes $g \rightleftharpoons q\bar{q}$ i.e gluons and $q\bar{q}$ pairs are already present. Fock states in all three cases are mentioned in the table given below. The table 2 differentiates the impact of presence of strange quark-anti-quark pair in sea in each case. The table 3 shows the probability contribution to sea in terms of different number of strange quark and gluon Fock states.

Case I II and III all shows the lower likelihood for the occupancy of Fock states where $s\bar{s}$ pair is present than those without $s\bar{s}$. The constraint to the available energy is being imposed by the coefficient and $C_{l-1} = \dfrac{2M_s}{M_B - sM_s - 2(l-1)M_s}$, $M_s$ is the mass of s-quark and $M_B$ is the mass of the baryon. In table 2, the probability enhances in all cases by almost 0 to 85% as we exclude $s\bar{s}$ pairs. In table 3, where we already have $s\bar{s}$ pair in addition to gluons, the relative probability for a single $s\bar{s}$ pair is large as compared to other combinations and it is least for $\left| s\bar{s}g \right\rangle$ possibly with a reason where a larger multiplicity leads to the reduced possibility. The results are shown in table 4, where magnetic moment and weak decay couplings are given on the basis of different assumption. Here Case I with strange sea favors the experimental values, there by suggesting the gluons in the sea to be generating strange quark condensates with similar quantum numbers. The data shifts by 7 to 46% approaching the experimental value favoring the strangeness in the sea. Here vector strange sea dominates and favors the results. Present framework suggests a stronger base to choose the model with suggested cases to verify the experimental and other theoretical values and hence provide a deeper understanding to the strange baryon structure.

| Sr. No | Fock State | Case-I | | Case-II | | Case-III | |
|---|---|---|---|---|---|---|---|
| | | With $s\bar{s}$ | Without $s\bar{s}$ | With $s\bar{s}$ | Without $s\bar{s}$ | With $s\bar{s}$ | Without $s\bar{s}$ |
| 1. | $|g\rangle$ | 0.029 | 0.165 | 0.029 | 0.165 | 0.111 | 0.103 |
| 2. | $|gg\rangle$ | 0.002 | 0.083 | 0.003 | 0.083 | 0.086 | 0.059 |
| 3. | $|ggg\rangle$ | 0.009 | 0.02 | 0.006 | 0.03 | 0.021 | 0.015 |
| 4. | $|\bar{u}u\rangle$ | 0.165 | 0.165 | 0.148 | 0.148 | 0.135 | 0.184 |
| 5. | $|\bar{d}d\rangle$ | 0.082 | 0.082 | 0.073 | 0.073 | 0.068 | 0.092 |
| 6. | $|\bar{u}ug\rangle$ | 0.023 | 0.072 | 0.023 | 0.072 | 0.149 | 0.203 |
| 7. | $|\bar{d}dg\rangle$ | 0.012 | 0.036 | 0.011 | 0.036 | 0.075 | 0.102 |
| 8. | $|\bar{u}ugg\rangle$ | 0.005 | 0.039 | 0.006 | 0.02 | 0.011 | 0.013 |
| 9. | $|\bar{d}dgg\rangle$ | 0.004 | 0.019 | 0.004 | 0.03 | 0.005 | 0.057 |

**Table 2: Fock States with Probabilities**

| Sr. No | Fock State | Case-I | Case-II | Case-III |
|---|---|---|---|---|
| 1. | $|\bar{s}s\rangle$ | 0.055 | 0.049 | 0.033 |
| 2. | $|\bar{s}sg\rangle$ | 0.002 | 0.002 | 0.050 |
| 3. | $|\bar{s}sgg\rangle$ | 0.042 | 0.028 | 0.027 |

**Table 3: Fock states with strange quark**

| Sr. No. | Property | Case-I | | Case-II | | Case-III | | Expt. Value[16] |
|---|---|---|---|---|---|---|---|---|
| 1. | $\mu_{\Xi^-}$ | -0.51 | -0.35 | -0.48 | -0.48 | -0.46 | -0.46 | -0.65±0.0025 |
| 2. | $g_A/g_{V(\Xi^-\to\Sigma^0)}$ | 0.89 | 0.80 | 0.82 | 0.82 | 0.79 | 0.72 | 1.29[17] |
| 3. | $g_A/g_{V(\Xi^-\to\Lambda)}$ | 0.25 | 0.19 | 0.24 | 0.24 | 0.23 | 0.21 | 0.25±0.5 |
| 4. | $g_A/g_{V(\Xi^-\to\Xi^0)}$ | -0.25 | -0.27 | -0.22 | -0.22 | -0.21 | -0.19 | 0.31[18] |

**Table 4: Static properties at low energy**